\documentclass[aps,pra,twocolumn,superscriptaddress]{revtex4-1}
\usepackage{graphicx}
\usepackage{hyperref}
\usepackage{amsmath}
\usepackage{amssymb}
\usepackage{epstopdf}
\usepackage{multirow}
\usepackage{threeparttable}
\usepackage{xcolor}
\usepackage{ulem}
\usepackage{natbib} 

\begin{document}

\title{Wide-field mid-infrared edge-enhanced upconversion imaging}
\author{Mengyao Yu}
\thanks{These authors contributed equally to this work.}
\affiliation{State Key Laboratory of Precision Spectroscopy, and Hainan Institute, East China Normal University, Shanghai 200062, China}

\author{Zhuohang Wei}
\thanks{These authors contributed equally to this work.}
\affiliation{State Key Laboratory of Precision Spectroscopy, and Hainan Institute, East China Normal University, Shanghai 200062, China}

\author{Jianan Fang}
\email{jnfang@lps.ecnu.edu.cn}
\affiliation{State Key Laboratory of Precision Spectroscopy, and Hainan Institute, East China Normal University, Shanghai 200062, China}
\affiliation{Chongqing Key Laboratory of Precision Optics, Chongqing Institute of East China Normal University, Chongqing 401121, China}

\author{Jixi Zhang}
\affiliation{State Key Laboratory of Precision Spectroscopy, and Hainan Institute, East China Normal University, Shanghai 200062, China}

\author{Tingting Zheng}
\affiliation{State Key Laboratory of Precision Spectroscopy, and Hainan Institute, East China Normal University, Shanghai 200062, China}

\author{Shina Liao}
\affiliation{State Key Laboratory of Precision Spectroscopy, and Hainan Institute, East China Normal University, Shanghai 200062, China}

\author{Kun Huang}
\email{khuang@lps.ecnu.edu.cn}
\affiliation{State Key Laboratory of Precision Spectroscopy, and Hainan Institute, East China Normal University, Shanghai 200062, China}
\affiliation{Chongqing Key Laboratory of Precision Optics, Chongqing Institute of East China Normal University, Chongqing 401121, China}
\affiliation{Collaborative Innovation Center of Extreme Optics, Shanxi University, Taiyuan, Shanxi 030006, China}

\author{Heping Zeng}
\affiliation{State Key Laboratory of Precision Spectroscopy, and Hainan Institute, East China Normal University, Shanghai 200062, China}
\affiliation{Chongqing Key Laboratory of Precision Optics, Chongqing Institute of East China Normal University, Chongqing 401121, China}
\affiliation{Shanghai Research Center for Quantum Sciences, Shanghai 201315, China}
\affiliation{Chongqing Institute for Brain and Intelligence, Guangyang Bay Laboratory, Chongqing, 400064, China}

\begin{abstract}
Edge-enhanced imaging is critical for visualizing weakly absorbing and transparent objects. Extending this functionality into the mid-infrared (MIR) region enables chemical sensitivity and improved imaging performance for biomedical, material, and remote-sensing applications. Here, we present a wide-field MIR edge-enhanced upconversion imaging system that integrates vortex-pump complex-amplitude engineering with aperiodic quasi-phase matching. In contrast to the bright-field modality, the wide-field edge-enhanced operation shows sensitive dependence on the crystal position relative to the Fourier plane. The system achieves single-shot operation with a 25-mm field of view and 79-$\mu$m spatial resolution, yielding a record-high space-bandwidth product of $7.9 \times 10^4$. We show that this capability enables direct visualization of phase gradients in transparent optical elements and enhances structural contrast in biological specimens. The demonstrated architecture combines high sensitivity, spectral specificity, and robust edge detection, offering a promising route toward advanced MIR imaging in industrial inspection and biomedical diagnostics.
\end{abstract}

\maketitle

\section{Introduction}
All-optical edge detection is a crucial image processing technique that enables high-speed acquisition of intensity or phase gradients of low-contrast objects, which has led to widespread applications in various fields, including label-free histopathology, machine vision, and industrial quality inspection \cite{Zhou2020NP}. Comparing to numerical processing methods, the optical analog approach has gained prominence for its ultrafast processing speed and ability to handle parallel image information efficiently \cite{Hu2024AdvPhoto, He2022Nanophoto}, thus addressing the significant challenges for conventional digital hardwares, particularly in terms of computation speed and energy efficiency. Moreover, the ability to acquire additional information from transparent objects, such as object phase and polarization, facilitates the observation of biological tissues and polarization-sensitive materials \cite{Wang2024NREE, Zhou2020SA}. Recently, various methods for optical edge detection have been implemented based on, for instance, high pass filtering to suppress low-frequency spatial components \cite{Cotrufo2023NC}, spiral phase mapping to perform Hilbert transformation \cite{Furhapter2005OE}, or wave-front shaping by micro-nano materials to directly modify the point spread function of the imaging system \cite{Liu2022NC, Qiu2025NC}. 

To date, all-optical edge-enhanced imaging predominantly operates within the visible or near-infrared regions due to the accessibility to relatively mature techniques or devices for light manipulation and detection. Nowadays, there is a significant impulse to extend the operation wavelength of edge-enhanced imaging into the mid-infrared (MIR) region \cite{Swartz2024SA}, pertaining to unique capabilities to reveal chemical and thermometric information and to resist light scattering in adverse conditions of dust, haze, and low-altitude clouds. However, highly sensitive room-temperature MIR imaging is challenging for narrow-bandgap or micro-bolometric sensors due to high dark current and severe thermal disturbance \cite{Rogalski2011IPT, Wang2019Small}. Notably, significant progresses have been made to elevate the operation temperature by adopting novel platforms based on low-dimensional materials \cite{Wu2025NC, Xue2024LSA, Long2017SA}, albeit with pressing difficulties in low light absorption, chemical instability, or complex manufacturing procedures. Therefore, it is highly demanded to develop novel techniques to realize high-sensitivity MIR edge-enhanced imaging at room temperature.

Alternatively, frequency upconversion strategy has emerged as an effective approach for sensitive and fast infrared imaging, where MIR signals are nonlinearly converted into the visible or near-infrared range to leverage high-performance silicon-based detectors \cite{Barh2019AOP, Mrejen2020LPR, Knez2022SA, Huang2022NC}. The indirect approach has led to superior imaging performances beyond the reach of conventional MIR focal plane arrays, particularly favoring high detection sensitivity at the single-photon level \cite{Huang2022NC, Dam2012NP, Paterova2020SA, Kviatkovsky2020SA} and fast frame rates at the high-definition megapixel format \cite{Junaid2019Optica, Zhao2023NC, Fang2024NC}. Moreover, the wave-mixing interaction during the parametric conversion allows amplitude and/or phase manipulation of the incident infrared signal via all-optical engineering of the complex pump field \cite{Manurkar2016Optica, Ansari2018Optica, Qiu2018Optica}. Such a nonlinear pump filter favors high-fidelity MIR optical modulation, which could circumvent parasitic diffraction effects at long wavelengths to improve the imaging contrast and spatial resolution \cite{Wang2023NC}. This unique feature has recently been exploited to implement nonlinear spatial filtering in spiral phase contrast (SPC) imaging \cite{Liu2019LP, Liu2019PRAppl, Junaid2020AO, Wang2021LPR, Zeng2023LPR, Gao2025OE}, where the required helical phase pattern is transferred from the vortex pump to the upconverted field at the Fourier plane within the nonlinear crystal. Moreover, the topological optical differentiation performance could be enhanced by performing complex-field pump filtering \cite{Yan2023OL}. Intriguingly, anisotropic \cite{Li2024OE} or curved \cite{Liu2020OE} edge enhancement is feasible by resorting to composite Laguerre-Gaussian or high-order vortex filters, respectively. Hence, the upconversion SPC imaging offers a sensitive and flexible way for the MIR edge-enhanced detection.
 
However, the involved intermediate conversion process typically requires stringent phase matching to facilitate a pronounced conversion efficiency, thus imposing a constraint on the field of view (FOV) of the upconversion imaging system \cite{Huang2022NC, Ge2023PRAppl}. Consequently, the acceptable incident angles are generally limited to just a few degrees, which is especially pronounced for long nonlinear crystals to enhance the interaction length \cite{Wang2021LPR}. The resulting space-bandwidth product (SBP), which quantifies the number of resolvable elements, is thus restricted by the trade-off between the spatial resolution and FOV. To increase the FOV, several methods have been proposed by translating the object plane \cite{Kehlet2015OE}, rotating the nonlinear crystal \cite{Junaid2019Optica}, or varying the crystal temperature \cite{Dam2012NP, Liu2019PRAppl}. These schemes usually necessitate parameter tuning and post-processing to obtain a stitched full-FOV image. Another solution is to employ broadband pump sources or polychromatic signal illumination to enlarge the acceptance angle, yet the wavelength-dependent magnification factor would inevitably result in imaging distortion \cite{Junaid2020AO, Fang2024NC, Ashik2019PR}. Very recently, single-shot realization of extended FOVs has been investigated through quasi-phase matching in chirped-poling crystals, albeit only demonstrated in the wide-field bright-field imaging modality \cite{Huang2022NC}. Notably, nonlinear metasurfaces \cite{Morales2021AP, Molina2024AdvMat} and thin films \cite{Manattayil2024LPR, Zhu2022Optica} provide a promising nanophotonic platform for phase-matching-free nonlinear conversion, yet at the cost of reduced conversion efficiency and detection sensitivity. Therefore, it remains an appealing quest to achieve wide-field MIR edge-enhanced imaging at high sensitivity and high resolution.

In this work, we have devised and implemented a wide-field edge-enhanced MIR upconversion imaging system based on synergic operations of aperiodic quasi-phase matching and nonlinear vortex filtering. In contrast to the bright-field instantiation, the SPC configuration needs precise alignment between the central singularity of the vortex pump and the principal axis of the imaging system, which has been addressed by additional efforts on light manipulation and mode matching. Specifically, the edge-enhanced performance is found to be sensitive to the crystal position relative to the Fourier plane, which allows one to observe the transition between the bright-field and dark-field imaging modalities. In the experiment, the acceptance angle is enlarged to about 28.1$^\circ$, corresponding to a FOV of 25 mm and spatial resolution of 79 $\mu$m. The resulting SBP reaches an unprecedented value of 7.9$\times 10^4$, which represents a significant landmark among reported MIR edge-enhanced imagers. Moreover, high-contrast visualization of real-world objects has been conducted to reveal the phase gradient in the transparent phase plate and to observe the fine structural details in biological samples. The integration of edge-enhanced and wide-field capabilities not only improves the image contrast for subtle structural details, but also enhances the image efficiency for high-throughput examination. In combination with chemical selectivity, the presented MIR imaging system would be useful in biomedical histopathology, material characterization, and environmental monitoring.

\begin{figure}[t!]
	\includegraphics[width=0.9 \columnwidth]{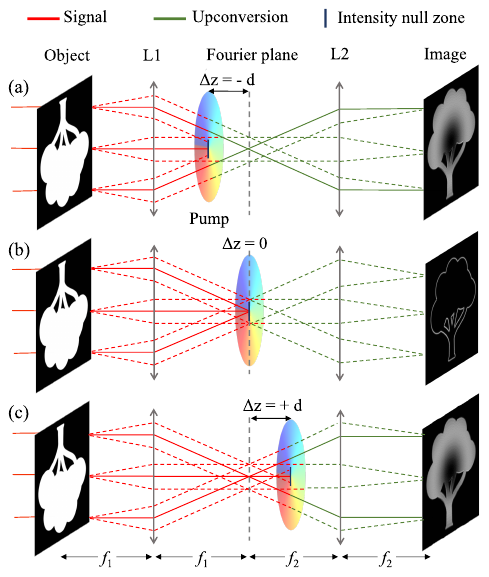}
	\caption{Concept for the position-dependent upconversion edge-enhanced imaging. The imaging system is arranged in the 4f configuration, where the nonlinear crystal is placed around the Fourier plane with a relative position of z. The vortex pump exhibits a donut intensity distribution and a helical phase pattern. The MIR signal from the illuminated object is focused into the crystal. The incident components with different spatial frequencies undergo nonlinear conversion, if not passing through the intensity null zone. (a-c) present the upconversion image at $\Delta z$ = -d, 0, +d, respectively. Wide-field edge-enhanced imaging can only be realized for $\Delta z$=0, where all low-frequency components of the object are fully blocked by the nonlinear vortex filter. Note that the crystal length is omitted here for simplicity, which actually corresponds to an effective range of involved poling periods during the conversion as detailed in Supplementary Note 3.}
	\label{fig1}
\end{figure}

\section{Basic principle}
The core mechanism of the MIR upconversion SPC imaging relies on all-optical modulation via the nonlinear parametric interaction \cite{Qiu2018Optica, Wang2021LPR}. Particularly, sum frequency generation (SFG) is commonly adopted due to high-efficiency second-order nonlinearity. Under the approximations of paraxial propagation and slowly varying envelope, the SFG process can be described by the coupled wave equation:
\begin{equation}
	\frac{\partial E_u}{\partial z}=\frac{2i\omega_u^2 d_\text{eff}}{k_u c^2} E_s E_p\exp^{-i \Delta kz} \ , 
	\label{eq1}
\end{equation}
where $E_{s,p,u}$ represent electric amplitudes of signal, pump, and upconverted fields, $d_\text{eff}$ is the effective nonlinear coefficient, and $c$ is the light speed in vacuum. The involved angular frequencies $\omega_{s,p,u}$ satisfy the energy conversion as  $\omega_{s} + \omega_{p} = \omega_{u}$.  And the momentum conservation is manifested by the phase-matching condition as $\Delta \vec{k} = \vec{k}_u - \vec{k}_s - \vec{k}_p - 2\pi/\Lambda$ = 0, where $\vec{k}_{s,p,u}$ are the wave vectors for the optical fields and $\Lambda$ is the poling period. Consequently, the SFG field would be simply related to the product of the signal and pump fields, which indicates that the amplitude and phase manipulation of the signal can be realized by engineering the complex field of the pump. The nonlinear wave mixing not only results in the frequency upconversion for sensitive infrared detection, but also provides an effective way to perform high-fidelity broadband MIR modulation.

In general, the wide-field phase matching necessitates the adaptation of operation parameters for various incident angles, for instance, resorting to tuning the crystal temperature \cite{Dam2012NP, Liu2019PRAppl} or tilting the crystal angle \cite{Junaid2019Optica}. Alternatively, chirped-poling nonlinear crystals can be used to facilitate a large acceptance angle, provided that the axial phase-mismatching term at an incident angle of $\theta_s$ is compensated by the position-dependent poling period as $\Delta k_z=k_u\cos \theta_u - k_s\cos \theta_s - k_p - 2\pi/\Lambda(z)$. The chirped quasi-phase-matching technique has been used to implement wide-field \cite{Huang2022NC} and broadband \cite{Mrejen2020LPR} upconversion bright-field imaging. However, the extension to edge-enhanced MIR upconversion imaging has yet been demonstrated, which calls for further investigation on wide-field vortex filtering performance.

 As shown in Fig. \ref{fig1}, the upconversion imaging system is usually constructed in the 4f configuration, where nonlinear conversion occurs at the Fourier plane \cite{Barh2019AOP}. In this case, the pump field acts as the transfer function, which modulates the amplitude and/or phase of the spatial frequency components of the object. Mathematically, the Fourier spectrum after the pump filtering operation of $E_p(\rho, \varphi)$ can be expressed as
\begin{equation}
\tilde{E}_f(\rho, \varphi) = \tilde{E}_s(\rho, \varphi) \times E_p(\rho, \varphi) \ .
\label{eq3}
\end{equation}
where $\tilde{E}_s(\rho, \varphi) = \mathcal{F}[E_s(r, \phi)]$ is the Fourier transform of the signal object image described by $E_s(r, \phi)$. The essence of the edge-enhancement operation lies in filtering out the low-frequency components, which can resort to either the high-pass filter via a hollow intensity distribution \cite{Junaid2020AO} or the interference destruction effect with a spiral phase pattern \cite{Liu2019LP, Liu2019PRAppl}. Here, a first-order Laguerre-Gaussian ($\rm{LG_{01}}$) pump is used to enhance the edge-enhancing performance \cite{Guo2006OL}. The complex amplitude distribution of the $\rm{LG_{01}}$ mode is given by 
\begin{equation}
	E_{p}(\rho,\varphi)=\frac{\rho}{\omega^2} e^{-(\frac{\rho}{\omega})^2} e^{i \varphi} \ ,
	\label{eq4}
\end{equation}
where $\omega$ is the beam radius. The LG-based vortex filter can lead to shaper edges comparing to the schemes based on solely amplitude or phase modulation, as discussed and compared in Supplementary Note 1. 

In contrast to the Gaussian pump, the vortex filtering performance depends on the relative distance $\Delta z$ away from the Fourier plane, which is ascribed to the inherent dark-field block of spatial frequency components. Specifically, wide-field edge enhancement can only be achieved when the optical vortex filter is placed at the Fourier plane, $\textit{i.e.},$ $\Delta z$ = 0, as shown in Fig. \ref{fig1}(b). In this case, low-frequency components for all the points in the object plane can be completely optically blocked by the intensity null zone of the LG beam. As the vortex filter deviates from its optimal position, the edge-enhancement effect gradually diminishes in the peripheral regions, where the lateral shift of the Fourier spectrum permits low-frequency components to pass through the vortex filter, as depicted in Figs. \ref{fig1}(a) and (c). In the extreme case with a large $\Delta z$, the edge-enhanced phenomenon almost disappears, and the imaging modality turns to the bright field. Therefore, the involved transition behavior could offer an effective way to acquire dual-mode image information, facilitating both global visualization of morphologic profile and detailed examination of fine structures.

\begin{figure*}[t!]
	\centering
	\includegraphics[width=0.80\textwidth]{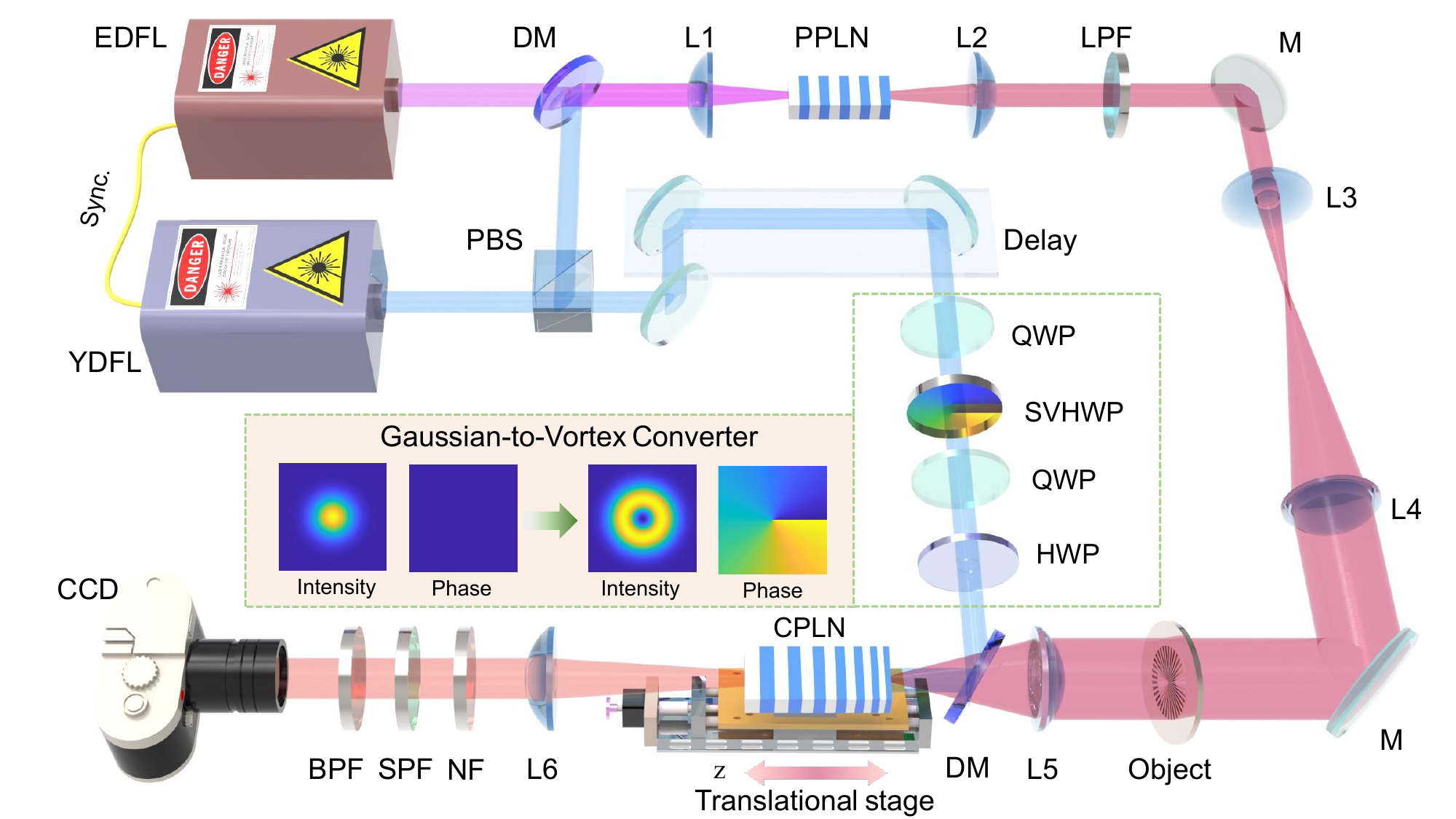}
	\caption{Experimental setup for the wide-field MIR edge-enhanced upconversion imaging system. The involved light sources originate from two synchronized erbium- and ytterbium-doped fiber lasers (EDFL and YDFL) at 1550 and 1030 nm, respectively. The dual-color beams are used to prepare the MIR signal at 3070 nm via different frequency generation in a periodically poled lithium niobate (PPLN) crystal. Then, the MIR beam is expanded to illuminate the object in a 4f imaging system, where a chirped-poling lithium niobate (CPLN) is used to perform the sum-frequency generation. A portion of the YDFL output serves as the pump after passing through a Gaussian-to-vortex converter. The CPLN is positioned on a translational stage to optimize the wide-field edge enhancement. Finally, the upconverted image is captured by a silicon camera. BS, beam splitters; PBS, polarization beam splitter; L, lens; DM, dichroic mirror; M, silver mirror; QWP and HWP, quarter- and half-wave plates; SVHWP, spatially varied HWP; BPF and SPF: band- and short-pass filter; NF: notch filter.}
	\label{fig2}
\end{figure*}

\section{Experimental section}
Figure \ref{fig2} illustrates the experimental setup for the wide-field MIR upconversion edge-enhanced imaging system. The whole system mainly consists of three parts, including the MIR signal generation, vortex pump  preparation, and wide-field upconversion imaging. The involved laser sources are from an ytterbium-doped fiber laser (YDFL) at 1030 nm and an erbium-doped fiber laser (EDFL) at 1550 nm. The two lasers are mode-locked at repetition rates around 14.6 MHz, which are passively synchronized through cross-phase modulation within an optical fiber \cite{Wang2021LPR}. The dual-color pulses are spatially combined by a dichroic mirror (DM) before being focused by an achromatic lens into a periodically poled lithium niobate (PPLN) crystal. Consequently, synchronous MIR pulse at 3070 nm can be generated through different frequency generation (DFG) process. The MIR beam is expanded to a diameter about 2.45 cm to provide coherent illumination for a transmission object. 

Meanwhile, the other part of YDFL output serves as the pump source, which is temporally controlled via a delay line and spatially manipulated by a Gaussian-to-vortex converter. The core of the mode converter is a  spatially varied half-wave plate (SVHWP, Thorlabs, WPV10L-1064). The SVHWP, often referred to as a q-plate, can generate a vortex beam by manipulating the polarization state of incident light and converting it into orbital angular momentum. The resulting pump beam exhibits a donut-shaped intensity distribution and a spiral phase pattern, closed to the LG$_{01}$ mode. It is worth noting that the pump is carefully controlled to maintain the same vertical polarization as the signal, which is required to facilitate the subsequent nonlinear conversion at the type-0 phase-matching condition. More details about the vortex beam generation and characterization are presented in Supplementary Note 4.

Finally, the pump and signal are mixed by another DM into a 4f imaging upconversion system based on the SFG process, where a chirped-poling lithium niobate (CPLN) is placed near the Fourier plane. The CPLN is fabricated with geometrical dimensions of 2$\times$3$\times$5 mm$^3$ (thickness $\times$ width $\times$ length), and has linearly ramping poling periods from 16 to 24 $\mu$m along the propagation direction. The chirped-poling design allows phase-matching adaptation for MIR signals at various incident angles, thus enlarging the FOV of the imaging system \cite{Huang2022NC}. The CPLN is placed on a translational stage, which favors precise optimization of edge-enhanced performance.  The upconverted SFG image at 771 nm passes through a series of spectral filters before being captured by a silicon-based electron-multiplying charge-coupled device (EMCCD, Andor iXonUltra 888).

\begin{figure*}[t]
	\centering
	\includegraphics[width=0.80\textwidth]{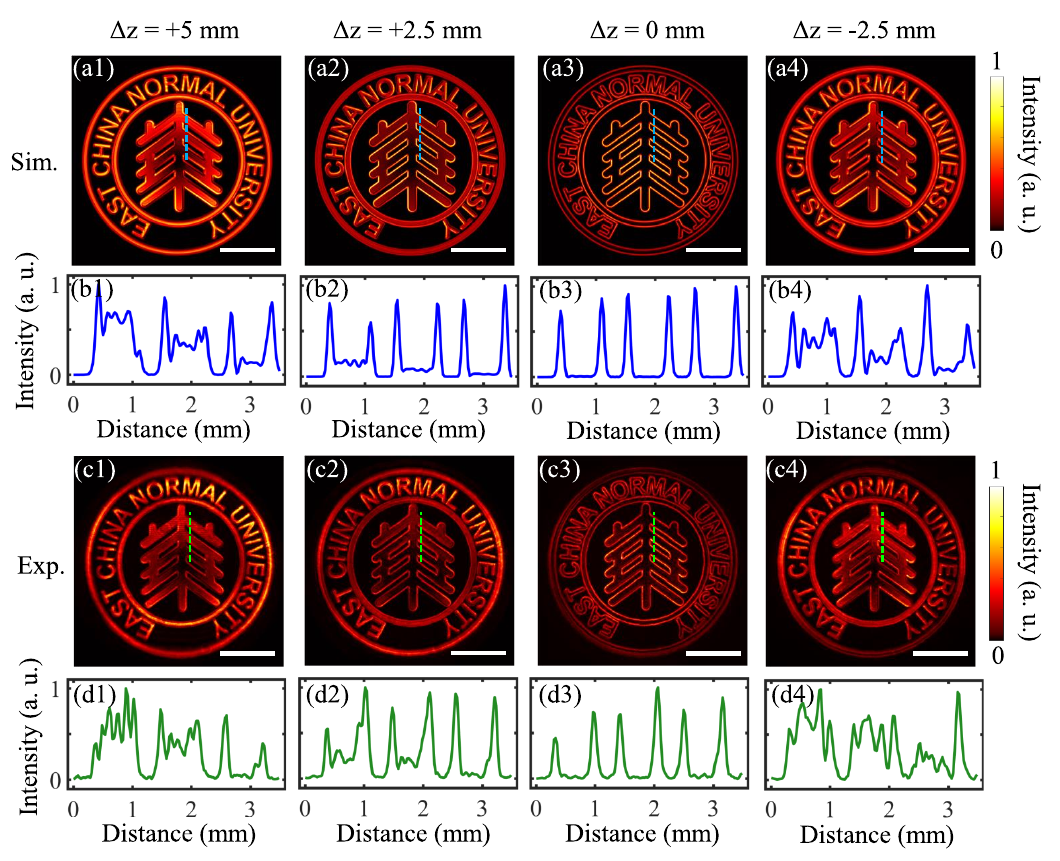}
	\caption{Numerical simulation (a1-a4) and experimental observation (c1-c4) for the vortex-pumped upconversion images as varying the position of the nonlinear crystal relative to the Fourier plane of the 4f imaging system. Corresponding cross sections are presented below the upconversion images at relative positions of $\Delta z$ = 5, 2.5, 0, and -2.5 mm, respectively. It can be seen that the wide-field edge-enhanced performance is optimized at $\Delta z$ = 0 mm. The scale bar is 4 mm. Note that the dynamic imaging behavior of varying the crystal position is recorded in Supplementary Video 1.}
	\label{fig3}
\end{figure*}

\section{Results and discussion}
\subsection{Wide-field nonlinear vortex filtering}
Now we begin with investigating the nonlinear vortex filtering performance for the wide-field upconversion imaging system. As discussed previously, the edge-detection operation under the LG-mode pumping relies both on high-pass filtering and spiral phase mapping for spatial-frequency components in the Fourier space. Consequently, the dark-field performance depends on the axial position of the nonlinear vortex filter with a reference to the Fourier plane. In the experiment, we have investigated the dependence by varying the nonlinear crystal along the light propagating direction. As discussed in Supplementary Note 3, the involved poling periods for a full acceptance angle of 28.1$^\circ$ range from 20.57 to 19.21 $\mu$m, corresponding to an effective interaction length of 0.82 mm. As a result, the nonlinear filtering operation is dominantly performed within a short active zone to achieve the wide-filed nonlinear conversion.

Figure \ref{fig3} presents the numerical simulation and experimental results for the upconversion images at various relative positions $\Delta z$ for the nonlinear crystal. As expected, the wide-field edge enhancement across the whole object can only be observed at $\Delta z$ = 0, where all the low-frequency components in the object plane are blocked by the optical vortex filter as shown in Fig. \ref{fig1}(b). As the relative position is tuned away from this ideal point, the edge enhancement starts to disappear at the periphery of the viewing field, which is consistent to the theoretical prediction in Figs. \ref{fig1}(a, c). Specifically, at a large deviation distance of $\Delta z$ = 5 mm, the edge-enhancing effect is only preserved in the small central area. The transition from the edge-enhanced to bright-field imaging modalities is clearly manifested in the cross-section traces, from both theoretical calculations in Figs. \ref{fig3}(b1-b4) and experimental observations in Figs. \ref{fig3}(d1-d4). We note the position-dependence vortex filtering effect is particularly pronounced at the wide-field operation because of the substantial off-axis displacement of the Fourier spectrum for outermost points in the object plane.

\begin{figure*}[t!]
	\centering
	\includegraphics[width=0.85 \textwidth]{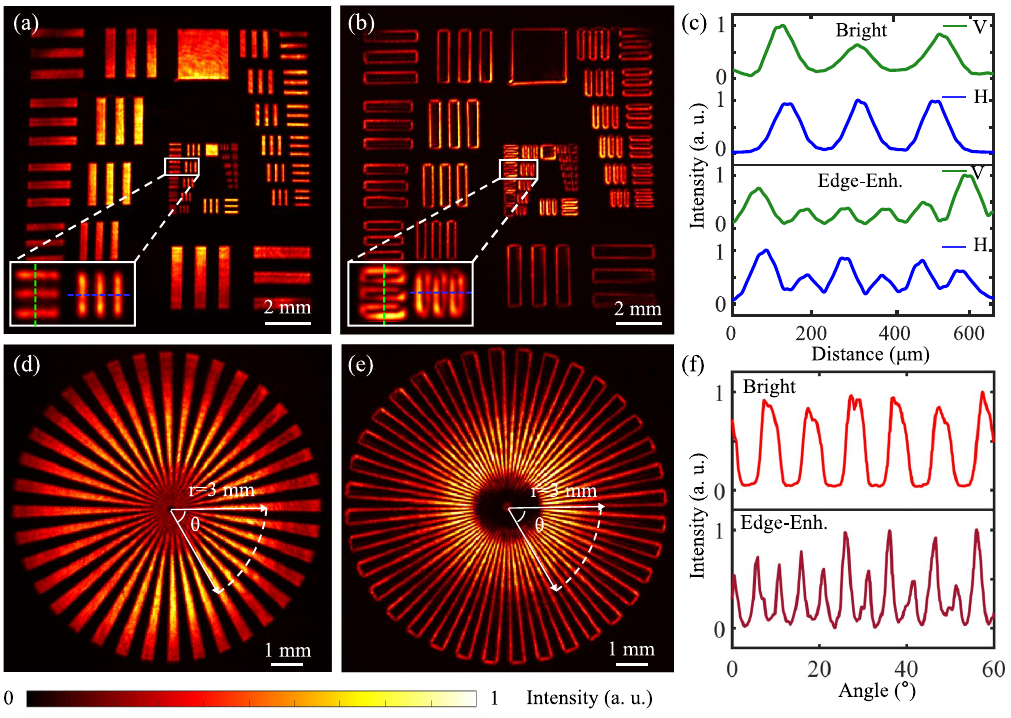}
	\caption{Wide-field MIR imaging performances at bright-field and edge-enhanced modalities. (a, d) Bright-filed images for the USAF resolution chart (a) and Siemens star test target (d). Corresponding edge-enhanced images are presented in (b) and (e), respectively. (c) Representative cross-section traces are given for the line pairs of third elements in the second group along the vertical and horizontal directions. (f) Cross sections at a radius of 3 mm. Note that the full FOV of the MIR imaging system can reach to about 25 mm, which is dictated by the one-inch optical elements in our experiment. And the spatial resolution is identified to be about 79 $\mu$m for edge-enhanced imaging, according to full width of maximum for the generated edge peak, as discussed in the text and Supplementary Note 5.}
	\label{fig4}
\end{figure*}

The observed sensitivity of wide-field edge-enhanced imaging performance to the crystal position is in marked contrast to the bright-field modality under the Gaussian pump. As shown in Supplementary Note 2, the wide-field bright-field imaging performance maintains over a large detuning range from -2.5 to 5 mm. The associated dynamic transitions for the edge-enhanced and bright-field imaging modalities are presented in Supplementary Video 1 for the sake of direct comparison. Moreover, we have also investigated the effect of the crystal temperature on the imaging performance. The edge-enhanced capability is almost kept at the operation temperatures from 25 to 125 $^\circ$C (see Supplementary Note 3 and Supplementary Video 2). Indeed, the phase-matched poling period varies about 0.31 $\mu$m. The corresponding position shift of the nonlinear vortex filter is as small as 0.19 mm, which plays a negligible role in affecting the imaging performance. The resulting robust operation eliminates the need to stabilize the crystal temperature as required in previous schemes \cite{Dam2012NP, Liu2019PRAppl}, thus improving the imaging stability and simplicity.

\subsection{Wide-field MIR edge-enhanced imaging}
Next, we turn to characterize the wide-field MIR edge-enhanced imaging performance. In the experiment, two resolution test targets are used to emulate the amplitude objects. Figures \ref{fig4}(a, b) give the measured upconversion images for a USAF 1951 resolution target at the bright-field and edge-enhanced modalities. The insets at the left-bottom corner present the zoom-in images for the third elements in the second group. The corresponding bar width is about 99.2 $\mu$m. The cross-section traces along the dashed lines are given in Fig. \ref{fig4}(c) in the bright-field modality. The contrasts of the obtained fringes are measured to be about 64\% and 85\% along the vertical and horizontal directions. The slightly better horizontal resolving capability is due to the asymmetric shape of the effective filtering aperture of the nonlinear crystal, $\textit{i.e.,}$ the width is longer than the thickness \cite{Huang2022NC}. Note that the spatial resolution can be further identified to be about 55.7 $\mu$m according to the Rayleigh criterion, where the contrast is defined as 11.1\% for rectangular shapes \cite{Barh2019AOP}. The cross sections for the edge-enhanced patterns are shown in Fig. \ref{fig4}(c), where the double peaks clearly reveal two edges of each bar, as expected for the first-order differential operation.

Similarly, a Siemens star test target with symmetric patterns is used to characterize the isotropic edge enhancement for the upconversion SPC imaging system. The bright-field and edge-enhanced images are presented in Figs. \ref{fig4}(d) and (e), respectively. A representative circular cross section at a radius of 3 mm is selected to show the edge-enhancing performance in Fig. \ref{fig4}(f). For the edge-enhanced imaging, the spatial resolution can be also identified by the Rayleigh criterion, where the edges are only detectable until the maximum point of one edge signal coincides with the first zero of another edge signal \cite{Liu2022NC}. In practice, one good estimation for the spatial resolution would be the linewidth of the measured edge, which is measured to be about 79 $\mu$m in our experiment as detailed in Supplementary Note 5. Additionally, the FOV of the presented imaging system is about 25 mm, mainly determined by the involved one-inch optical elements (see Supplementary Note 5 for the measured full FOV). As a result, the SBP can be calculated to be about 7.9$\times$10$^4$, which is about four-fold improvement over previously achieved values \cite{Junaid2020AO, Liu2019PRAppl, Zeng2023LPR, Wang2021LPR}. In Supplementary Note 6, a comparison table is presented to summarize the imaging performance for reported edge-enhanced upconversion imaging systems.

\begin{figure}[t!]
	\centering
	\includegraphics[width=0.95\columnwidth]{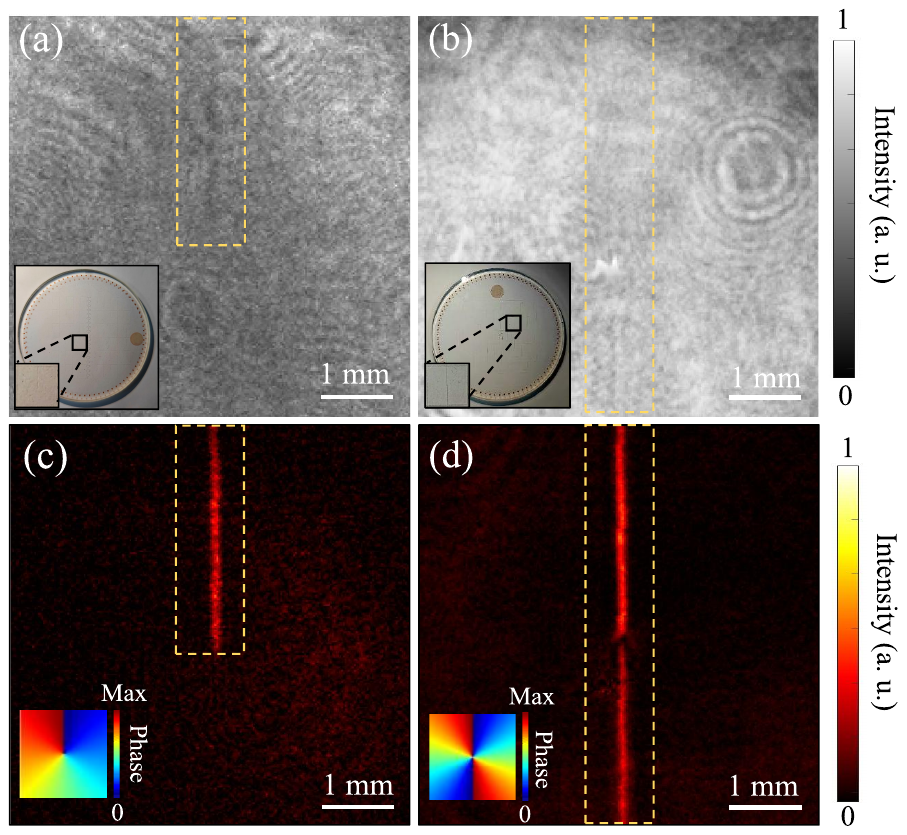}
	\caption{Spatial differentiation for phase gradient information. (a, b) Bright-field images of first-order (a) and second-order (b) spiral phase plates. The phase-discontinuous lines are highlighted by the dashed rectangles. Insets at the left-bottom corner present the photographs of the phase plates, where the imaging areas for selective elements are denoted by the black boxes. (c, d) Spatially differentiated images for the first-order (c) and second-order (d) vortex plates, which clearly reveal the phase-discontinuous lines. Insets show the corresponding helical phase patterns.}
	\label{fig5}
\end{figure}

\subsection{Edge-enhanced visualization for phase objects}
Furthermore, we have studied the edge-enhancement imaging performance in high-contrast visualization of weakly absorbed samples or phase objects. Indeed, the LG-mode filtering of spatial frequency components is analogous to the space-differentiation operation, which allows the acquisition of phase information for the interrogated amplitude-transparent sample \cite{Qiu2018Optica, Yan2023OL}. For instance, in the presence of an object with a unitary amplitude and a phase distribution of $e^{i \phi(x,y)}$, the spatial differentiation operation leads to output electric field of $ i e^{i \phi(x,y)} [\partial \phi(x,y) / \partial x]$ \cite{Liu2022NC}. The corresponding intensity distribution is thus proportional to phase gradients in the specimen, which would be useful to identify the morphological structure in material inspection and biological examination. 

As an example, Fig. \ref{fig5} presents the captured images for the vortex phase plates (VPPs) that are commonly used phase-type diffractive optical elements to generate an optical vortex with orbital angular momentum. The phase introduced by the VPP varies azimuthally around its center. The associated spiral phase pattern typically follows the form of $l \phi$, where $l$ is the topological charge, representing the number of 2$\pi$ phase windings around the optical axis. The bright-field images for first-order and second-order VPPs are given in Figs. \ref{fig5}(a, b), which exhibit nearly homogeneous intensity distribution as expected for a phase object. In contrast, the corresponding dark-field images in Figs. \ref{fig5}(c, d) clearly indicate the phase discontinuity lines because of the existence of the phase gradient across this boundary. 

\begin{figure}[t!]
	\centering
	\includegraphics[width=0.95\columnwidth]{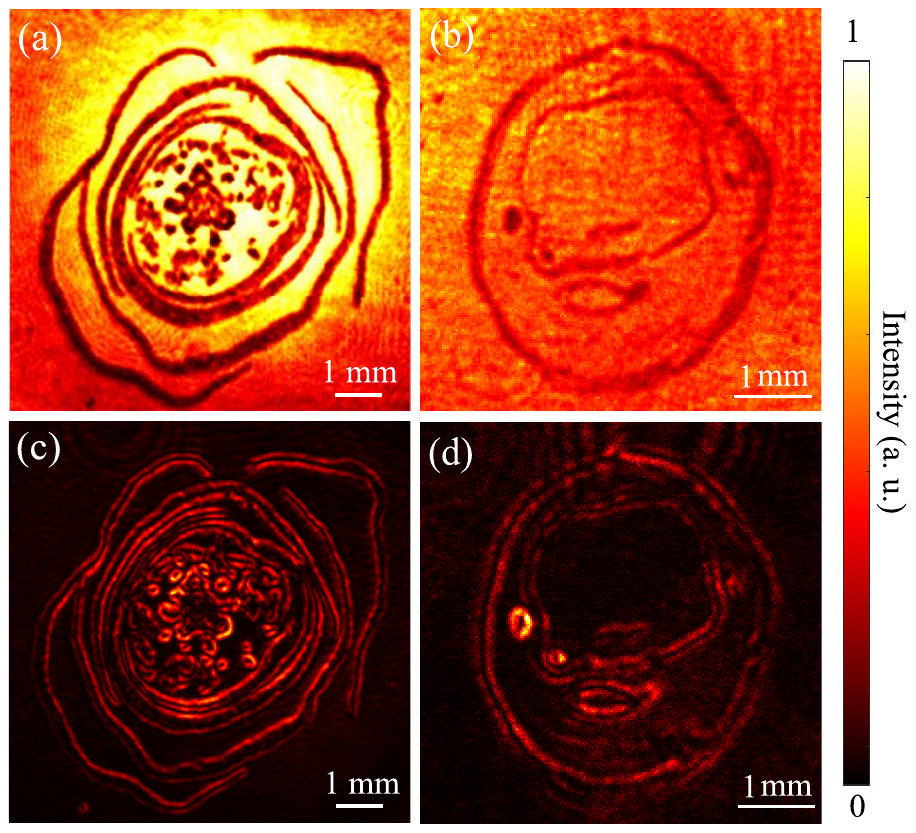}
	\caption{Edge-enhanced observation of biological specimens. (a, b) Bright-field images for sliced samples of \textit{Datura metel} L. (a) and \textit{Lumbricus terrestris} (b). (c, d) Corresponding spatially differentiated images, which show significant edge enhancement and high-contrast boundaries.}
	\label{fig6}
\end{figure}

Finally, real-world phase samples are used to illustrate the advantageous edge-enhancing feature to reveal fine structure details that are otherwise usually hidden in bright-field imaging modality. These specimens only significantly affect the phase, not its amplitude, and are thus referred to as phase objects \cite{Huo2024NC}. Figure \ref{fig6}(a) presents the bright-field image of a cross-sectional specimen of a \textit{Datura metel} L., which shows the structural boundaries of stamens and petals. The details about the stamens are highlighted in the edge-enhanced image as shown in Fig. \ref{fig6}(c). Similarly, the captured bright-field image of a \textit{Lumbricus terrestris} in Fig. \ref{fig6}(b) presents a low-contrast profile for the elements on the sample. From the edge-enhanced image in Fig. \ref{fig6}(d), we can distinctly resolve the body wall structure, the digestive system centered around the intestinal tract, and blood vessels. Therefore, the shapes and boundaries of fine structures in the two specimens are not very clearly discernible in the bright-field images due to their transparent nature. In contrast, the differentiated images obtained in the dark-field modality show both significant edge enhancement and high-contrast boundaries. Thanks to the high detection sensitivity, the dark-field MIR imaging is performed at an illumination density of only 0.3 $\text{mW} \ \text{cm}^{-2}$. The minimized phototoxicity favors non-destructive observation of biological specimens \cite{Junaid2019Optica}.

We note that the MIR illumination will be especially beneficial to penetration imaging for defect inspection through semiconductor or polymers materials \cite{Huang2022NC, Knez2022SA}. For example, germanium wafers widely used in fabricating MIR photonic chips are transparent for wavelength above approximately 2 $\mu$m. Additionally, the inherent chemical specificity in molecular spectroscopic analysis makes MIR imaging valuable in label-free examination of biomedical samples \cite{Junaid2019Optica, Zhao2023NC}, which would offer a powerful tool to reveal subtle pathological features by combining the edge enhancing and spectral resolving capabilities. 
 
\section{Conclusion}
In summary, we propose and demonstrate a wide-field MIR edge-enhancement upconversion imaging based on phase-matching engineering and pump field manipulation. The chirped-poling structure of the nonlinear crystal allows significant enlargement of the acceptance angle up to 28.1$^\circ$, about an order of magnitude larger than that in the single-period configuration. The wide-field operation of edge enhancement is found to be sensitive to the operation position of nonlinear vortex filtering along the propagation direction, which contrasts to the bright-field imaging under the Gaussian pump. The careful axial optimization of the spatial frequency filtration enables one to obtain a wide FOV with diameter of about  25 mm along with a spatial resolution of 79 $\mu$m. The corresponding SBP reaches to 7.9$\times$10$^4$, which represents a significant landmark among reported upconversion imaging schemes as shown in the comparison table  in Supplementary Note 6. Moreover, high-contrast visualization of phase objects has been demonstrated with vortex plates and biological samples, which highlights the unique feature of spatial differentiation in identifying critical shapes and boundaries for transparent specimens.

To go beyond the achieved performance, there are several aspects that deserve further investigation. First, the current spatial resolution is restricted by the crystal thickness, which would impose a limitation on the bandwidth of the spatial frequency in the Fourier plane. A practical route to further enhance the resolution is to adopt nonlinear Fourier ptychography \cite{Zheng2024Optica}, in which a sequence of upconversion images obtained under rotationally varied elliptical-pump filtering is computationally synthesized to expand the effective numerical aperture of the imaging system. Second, spectral imaging is possible by tuning the central wavelength of the MIR illumination \cite{Huang2022NC, Zhao2023NC} or adopting a broadband supercontinuum \cite{Fang2024NC}. The CPLN crystal used in the experiment has been designed to cover a wide spectral window from 2.5 to 5.0 $\mu$m, which accommodates absorption peaks associated with chemical bonds like C-H, C=O, N-H, and O-H \cite{Paterova2020SA, Kviatkovsky2020SA}. Third, the nonlinear optical modulation provides an effective approach to realize broadband MIR modulation at high efficiency and high fidelity, which can be readily extended to high-order optical spatial differentiation for further enhancing the sharpening effect on the edges of the sample \cite{Huo2024NC}. Additionally, flexible switching between bright-field and dark-field imaging modalities is feasible by alternating the pump mode, which facilitates on-demand acquisition of overall morphology and detailed information of the targeted specimen \cite{Zeng2025APL}.

\section*{Funding}
This work was funded by Shanghai Pilot Program for Basic Research (TQ20220104); National Natural Science Foundation of China (62175064, 62235019, 62035005); Innovation Program for Quantum Science and Technology (2023ZD0301000); Shanghai Municipal Science and Technology Major Project (2019SHZDZX01); Natural Science Foundation of Chongqing (CSTB2025NSCQ-GPX0443); Postdoctoral Fellowship Program and China Postdoctoral Science Foundation (GZC20250545, 2024M760918, 2025T180224); Fundamental Research Funds for the Central Universities (YBNLTS2025-009).

\section*{NOTES}
The authors declare no competing financial interest.

\section*{Supporting Information}
The Supporting Information is available free of charge at  https://pubs.acs.org/doi/XXXX.

\textbf{Supplementary Information:} Coupled wave equation model; Comparison to bright-field upconversion imaging; Temperature dependence performance; Vortex pump preparation; Spatial banwidth product estimation; Performance comparison table. (PDF)

\textbf{Supplementary Video 1:} Recorded dynamic behavior for the wide-field upconversion imaging under the Gaussian and vortex pump beams as varying the crystal position relative to the Fourier plane. The relative position is tuned from - 3 to 3 mm, with a step of 0.25 mm. (MP4)

\textbf{Supplementary Video 2:} Recorded edge-enhanced upconversion images as changing the operation temperature of the nonlinear crystal from 25 to 125 $^\circ\text{C}$ with an interval of 5 $^\circ\text{C}$. (MP4)

\section*{Data Availability Statement}
The data that support the findings of this study are available from the corresponding author upon reasonable request.

\section*{Keywords}
edge enhancement, mid-infrared imaging, spiral phase imaging, frequency upconversion, optical modulation, vortex beam

\end{document}